\documentclass[aps,prl,twocolumn,amsmath,amssymb,showpacs,superscriptaddress,notitlepage]{revtex4-1}
\usepackage[colorlinks=true,linkcolor=blue,anchorcolor=red,citecolor=blue, urlcolor=blue]{hyperref}
\usepackage{graphicx}
\usepackage{color}
\usepackage{bm}
\usepackage{float}
\usepackage{changes}

\colorlet{Changes@Color}{red}
\begin{document}

\title{Field-Tunable One-Sided Higher-Order Topological Hinge States in Dirac Semimetals}

\author{Rui Chen}
\affiliation{Shenzhen Institute for Quantum Science and Engineering and Department of Physics, Southern University of Science and Technology (SUSTech), Shenzhen 518055, China}
\affiliation{Shenzhen Key Laboratory of Quantum Science and Engineering, Shenzhen 518055, China}
\affiliation{School of Physics, Southeast University, Nanjing 211189, China}

\author{Tianyu Liu}
\affiliation{Max-Planck-Institut f\"ur Physik komplexer Systeme, 01187 Dresden, Germany}
\affiliation{Shenzhen Institute for Quantum Science and Engineering and Department of Physics, Southern University of Science and Technology (SUSTech), Shenzhen 518055, China}

\author{C. M. Wang}
\affiliation{Department of Physics, Shanghai Normal University, Shanghai 200234, China}
\affiliation{Shenzhen Institute for Quantum Science and Engineering and Department of Physics, Southern University of Science and Technology (SUSTech), Shenzhen 518055, China}
\affiliation{Shenzhen Key Laboratory of Quantum Science and Engineering, Shenzhen 518055, China}

\author{Hai-Zhou Lu}
\email{Corresponding author: luhz@sustech.edu.cn}
\affiliation{Shenzhen Institute for Quantum Science and Engineering and Department of Physics, Southern University of Science and Technology (SUSTech), Shenzhen 518055, China}
\affiliation{Shenzhen Key Laboratory of Quantum Science and Engineering, Shenzhen 518055, China}

\author{X. C. Xie}
\affiliation{International Center for Quantum Materials, School of Physics, Peking University, Beijing 100871, China}
\affiliation{Beijing Academy of Quantum Information Sciences, Beijing 100193, China}
\affiliation{CAS Center for Excellence in Topological Quantum Computation, University of Chinese Academy of Sciences, Beijing 100190, China}

\begin{abstract}
Recently, higher-order topological matter and 3D quantum Hall effects have attracted a great amount of attention.
The Fermi-arc mechanism of the 3D quantum Hall effect proposed to exist in Weyl semimetals is characterized by the one-sided hinge states, which do not exist in all the previous quantum Hall systems, and more importantly, pose a realistic example of the higher-order topological matter.
The experimental effort so far is in the Dirac semimetal Cd$_3$As$_2$, where however, time-reversal symmetry leads to hinge states on both sides of the top and bottom surfaces, instead of the aspired one-sided hinge states. We propose that under a tilted magnetic field, the hinge states in Cd$_3$As$_2$-like Dirac semimetals can be one sided, highly tunable by field direction and Fermi energy, and robust against weak disorder. Furthermore, we propose a scanning tunneling Hall measurement to detect the one-sided hinge states. Our results will be insightful for exploring not only the quantum Hall effects beyond two dimensions, but also other higher-order topological insulators in the future.
\end{abstract}

\maketitle

\begin{figure}[H]
\centering
\includegraphics[width=0.95\columnwidth]{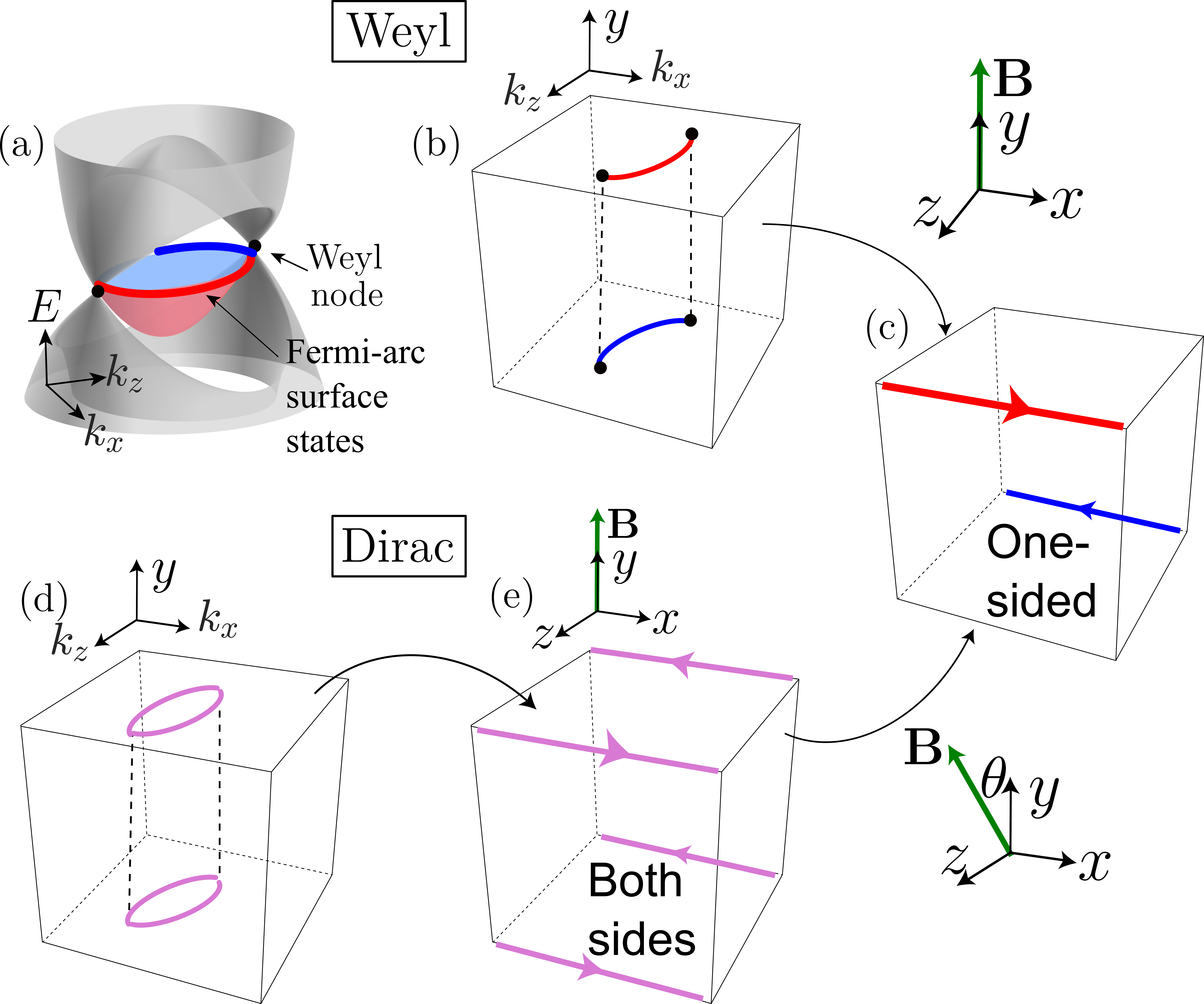}
\caption{(a) Energy dispersions of the bulk (gray, $k_y=0$) and Fermi-arc surface states [red for top and blue for bottom in (b)] of a Weyl semimetal. (b) The Fermi-arc surface states on the top and bottom surfaces can form a 2D electron gas distributed in the 3D system to support a 3D quantum Hall effect under a magnetic field $\mathbf{B}$. The 3D quantum Hall effect is characterized by the one-sided hinge states (red and blue arrows) in (c), which do not exist in all the previous quantum Hall systems. In the experimentally accessible Dirac semimetals, however, time-reversal symmetry leads to a complete 2D electron gas on each surface [(d)] and hinge states on both sides [(e)], instead of the desired one-sided hinge states. We find that the hinge states
in the Dirac semimetal can be one sided, highly tunable by a tilted magnetic field [from (e) to (c)] and Fermi energy [Fig.~\ref{fig_spectrum}], giving a signature of the Fermi-arc 3D quantum Hall effect and higher-order topological states of matter. There are no hinge states along the $x$ direction because of the assumed translational symmetry. }
\label{fig_illustration}
\end{figure}

{\color{red}\emph{Introduction}} -
Since the discovery of the quantum Hall effect~\cite{Klitzing80prl,Thouless82prl},
tremendous efforts have been devoted to the search for a quantum Hall effect beyond two dimensions \cite{Halperin87jjap,Montambaux90prb,Kohmoto92prb,
Koshino01prl,Bernevig07prl,Stormer86prl,Cooper89prl,
Hannahs89prl,Hill98prb,Cao12prl,Masuda16sa,Liu16nc,LiuJY19arXiv,Tang19nat,Qin20prl,LiH20prl,
ChengSG20prb,Wang20prbrc,Ma20arxiv,
ZhangSC2001science,Lohse2018Nature,Zilberberg2018Nature,PricePRL2015,Jin18prbrc,Molina2018PRL,Benito2020PRB,Chang2021PRB,Chang2021PRB1}. Recently, a 3D quantum Hall effect in
topological semimetals~\cite{WangCM17prl,Lu2018NatSciRev} has attracted a great amount of attention, in which the Fermi arcs from opposite surfaces and the Weyl nodes can form a ``Weyl orbit''~\cite{Potter14nc,ZhangY16srep,Moll16nat} to support the cyclotron motion of electrons driven by magnetic fields and a quantum Hall effect in 3D~[Figs.~\ref{fig_illustration}(a)-\ref{fig_illustration}(c)].
More intriguingly, this 3D quantum Hall effect is characterized by the one-sided hinge states, that is, the edge states of the Landau levels lying at one side of the top surface but at the opposite side of the bottom surface [Fig.~\ref{fig_illustration}(c)].
The one-sided hinge states
do not exist in all the previous quantum Hall systems and
can be regarded as an example of the long-sought
higher-order topological states of matter~\cite{Benalcazar2017Science,Langbehn2017PRL,Benalcazar2017PRB,Song2017PRL, Ezawa2018PRL,Ezawa2018PRL1,Schindler2018NatPhys,Serra_Garcia2018Nature,
Peterson2018Nature,Noh2018NatPho,Imhof2018NatPhys,Liu2019PRL,Roy2019PRB,Roy2019PRR,Serra_Garcia2019PRB,ChenR2020PRL,Hua2020PRB,LiuZR2021arXiv,
Queiroz2019PRL,Sitte2012PRL,ZhangFan13PRL,Yuria19PRL,Schindler2018SciAdv,Yan2018PRL,Yan2019PRL,
Varjas2019PRL,Choi2020NatMat,Ghorashi2020PRL,Wang2020PRL,Wieder2020NatCom,WangKai2020PRL,
Francesco2020PRR,Constantinos2019PRB,Frank2020PRB,Frank2020PRL,LiuVincent2020PRL1,LiuVincent2020PRL2,Xie2021NatRevPhy,LiCZ2020PRL}, where a $d$-dimensional
system can host $\left(d-2\right)$- or lower-dimensional boundary states protected by both topology and symmetry.
The experimental detection of higher-order topological insulators is a challenging problem,
because of demanding symmetry requirements.
In contrast, the one-sided hinge states in topological semimetals
not only may facilitate the detection of the higher-order topological states of matter but also can justify the 3D quantum Hall effect. Nevertheless, the quantum Hall effect has been observed only in slabs of the topological semimetal Cd$_3$As$_2$, with its mechanism still under debate~\cite{ZhangC17nc-QHE,Uchida17nc,Schumann18prl,Galletti2018PRB,Nishihaya2018SciAdv,Goyal2018APLMat,ZhangC19nat,Wangshuo2018PRL,Nishihaya2019NatCom,Lin2019PRL,
Yang2019NC,Kealhofer2020PRX}.
Moreover, Cd$_3$As$_2$ is a Dirac semimetal composed of two time-reversed Weyl semimetals, and time-reversal symmetry leads to hinge states on both sides of the top and bottom surfaces~[Figs.\ref{fig_illustration}(d)-\ref{fig_illustration}(e)], instead of the wanted one-sided hinge states~\cite{Wang13prb,Liang15nmat,Liu14nmat,Borisenko14prl}.

\onecolumngrid
\begin{figure*}[t]
\centering
\includegraphics[width=2\columnwidth]{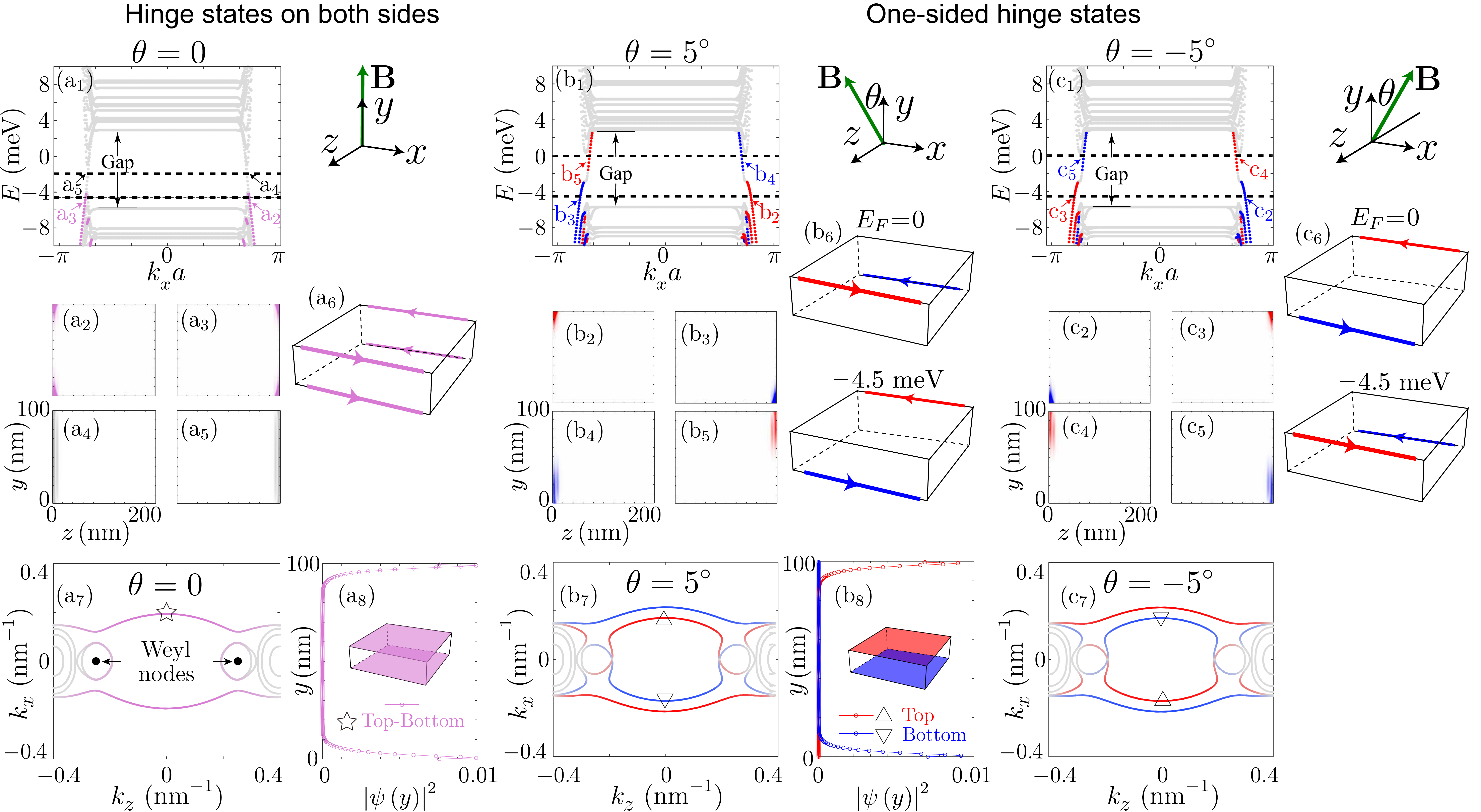}
\caption{(a$_1$) Energy spectrum of the Dirac semimetal ribbon in a magnetic field $B_y=15$ T ($\theta=0$), with open (periodic) boundary conditions along the $y$ and $z$ ($x$) directions. The ribbon size is $n_y a=100$ nm and $n_z a=200$ nm. The lattice constant $a=1$ nm. [(a$_2$)-(a$_5$)] The wave function distributions for the states marked by the horizontal dash lines in (a$_1$). Inside the Landau-level gap (at $E_F=-4.5$ meV), the hinge states lie on both sides, as schematically illustrated in (a$_6$).
(a$_7$) The Fermi surface of the Dirac semimetal slab in (a$_1$), when considering only the Zeeman effect of the magnetic field. (a$_8$) The wave function distribution of the state marked by the star in (a$_7$).
[(b$_1$)-(b$_8$)] The same as (a$_1$)-(a$_8$), except for $\theta=5^\circ$ ($B_z=B_y\tan\theta$). The hinge states become one sided at $E_F=0$ and switch their positions at $E_F=-4.5$ meV.
[(c$_1$)-(c$_7$)] The same as (a$_1$)-(a$_7$) except for $\theta=-5^\circ$. The one-sided hinge states are swapped compared to those in (b$_1$)-(b$_7$). Red, blue, and purple represent top, bottom, and top-bottom surfaces, respectively.}
\label{fig_spectrum}
\end{figure*}
\twocolumngrid

\begin{figure}[htbp]
\centering
\includegraphics[width=0.95\columnwidth]{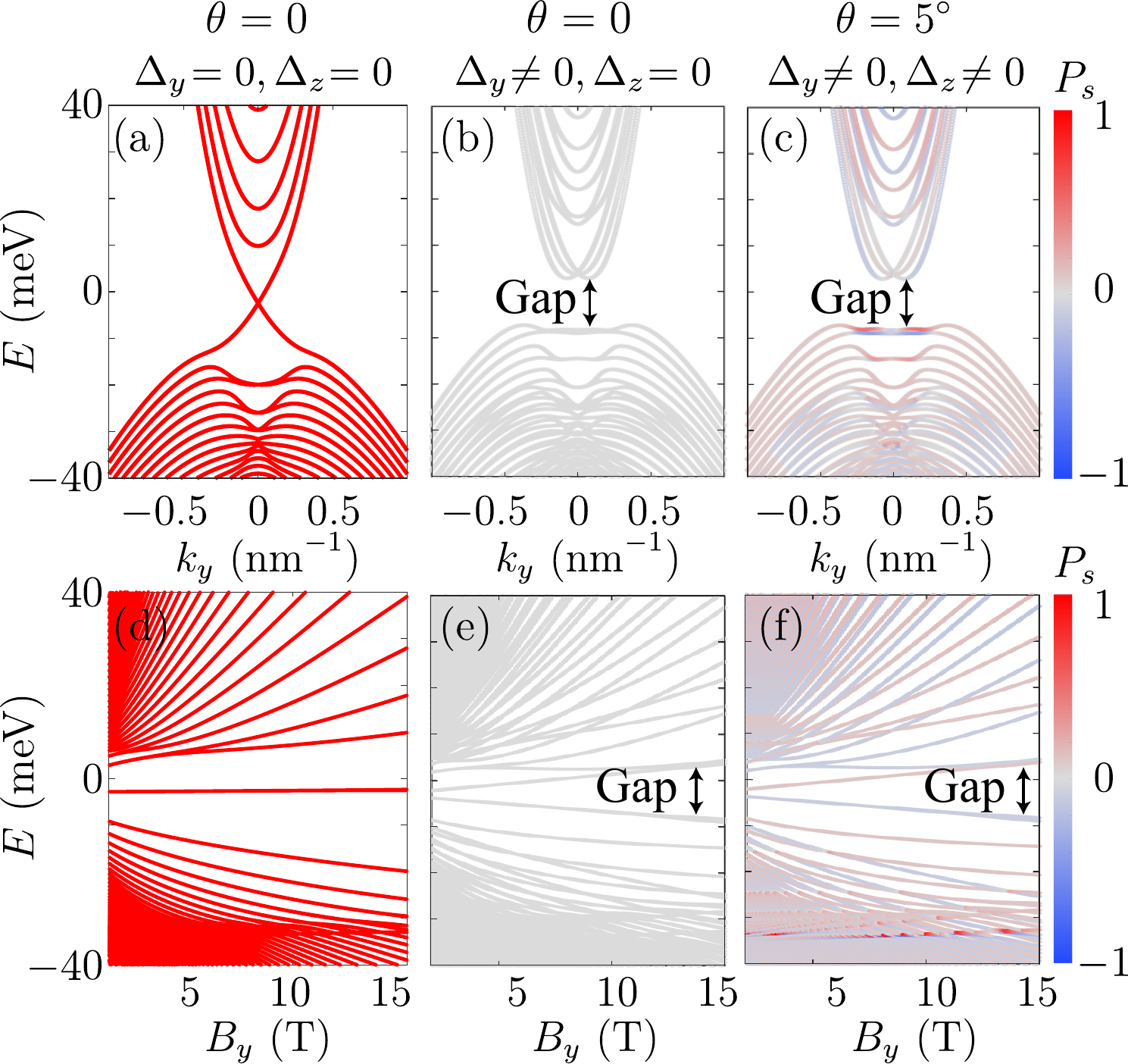}
\caption{The Landau bands of the Dirac semimetal (a-c) as functions of the wave vector $k_y$ at $B_y=15$ T and (d-f) as functions of the magnetic strength $B_y$ at $k_y=0$. The color bars indicate the weights of $h_0(\mathbf{k})$ (red) and $h_0^*(-\mathbf{k})$ (blue) in Eq. (\ref{Hamiltonian}). [(a), (d)] $\theta=0$ and the Zeeman coupling $\Delta_y=0$. Only the energy spectra for $h_0(\mathbf{k})$ are shown as $h_0(\mathbf{k})$ and $h_0^*(-\mathbf{k})$ show the same behavior. [(b), (e)] $\theta=0$ and $\Delta_y\neq 0$. [(c), (f)] $\theta=5^\circ$ and the Zeeman energies $\Delta_y\neq 0$ and $\Delta_z\neq0$. The orbital effect of $B_z$ is not included.}
\label{fig_Landaulevel}
\end{figure}

\begin{figure}[h!]
\centering
\includegraphics[width=0.95\columnwidth]{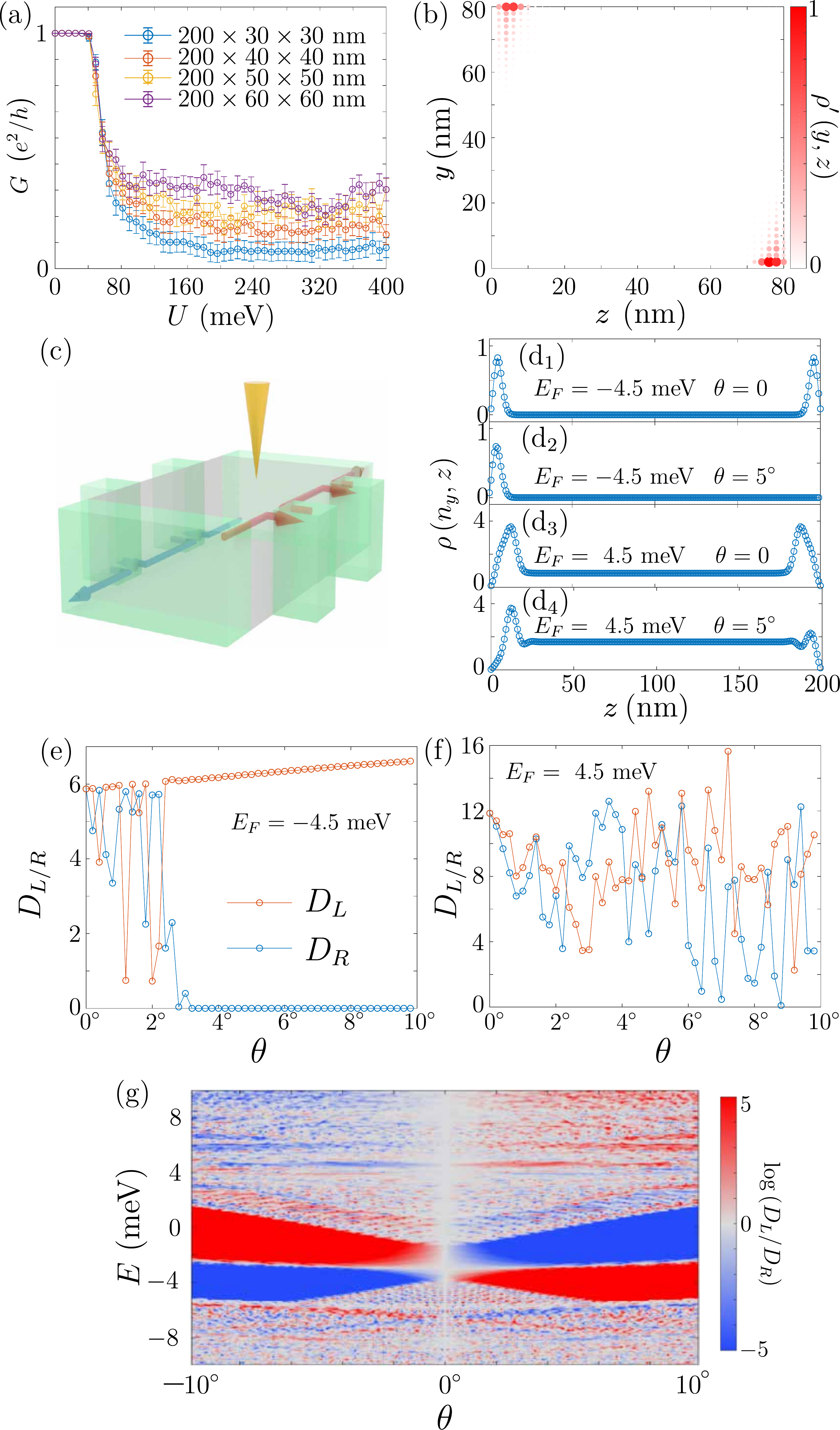}
\caption{(a) The disorder-averaged conductance as a function of the disorder strength $U$ for different system sizes. The error bars show the standard deviation for 200 samples. $E_F=-4.5$ meV. (b) The disorder-averaged local density of states with system size $n_x a=200$ nm, $n_y a=n_za=80$ nm. $U=20$ meV and $E_F=-4.5$ meV. (c) Schematic of the scanning tunneling Hall measurements for probing the hinge states. Gray, green, and yellow correspond to the Dirac semimetal, in-plane Hall-bar electrodes, and scanning tunneling microscopy tip, respectively. [(d$_1$), (d$_2$)] The local density of states $\rho\left(n_y,z\right)$ as a function of $z$ at $E_F=-4.5$ meV for $\theta=0$ and $\theta=5^\circ$, respectively. (e) The total local density of states for the 20 outermost layers near the left (L) and right (R) hinges on the topmost layer $D_{L/R}$ as functions of the tilting angle $\theta$ with $E_F=-4.5$ meV. [(d$_3$), (d$_4$), (f)] The same as (d$_1$), (d$_2$), and (e), except for $E_F=4.5$ meV.  (g) $\log(D_L/D_R)$ as a function of $E_F$ and $\theta$. In (c)-(g), the ribbon size is $n_y a=100$ nm and $n_z a=200$ nm; the lattice constant $a=1$ nm. }
\label{fig_disorder}
\end{figure}

In this Letter, we propose to realize the one-sided hinge states in Cd$_3$As$_2$-like Dirac semimetals in a tilted magnetic field [from Figs.~\ref{fig_illustration}(e) to \ref{fig_illustration}(c)].
With the help of a realistic model, we numerically calculate the energy spectra and the wave function distributions of Cd$_{3}$As$_{2}$ slabs grown along the experimentally accessible [110] crystallographic direction (Figs.~\ref{fig_spectrum} and \ref{fig_Landaulevel}), which show
the desired one-sided hinge states, with their positions tunable by the angle of the magnetic field $\theta$ and Fermi energy $E_F$.
We numerically check the robustness of the one-sided hinge states against weak disorder, and expect the one-sided hinge states to be detectable in the state-of-the-art scanning tunneling Hall measurement (Fig.~\ref{fig_disorder})~\cite{Schwenk2020RSI}. This work will not only help identify the Fermi-arc 3D quantum Hall effect but also provide insights for searching for higher-order topological states of matter.

{\color{red}\emph{Model}} -
We start with an effective Hamiltonian
of the [110]-Dirac semimetal Cd$_{3}$As$_{2}$ slab~\cite{Wang13prb,Wang2016PRBCdAs}, and consider both the orbital effect (Landau levels) and Zeeman effect of a magnetic field $\mathbf{B}=\left( 0,B_{y},B_z=B_y\tan \theta\right)$~\cite{LiH20prl}. Using the Landau gauge, the vector potential of the magnetic field $\mathbf{A}=\left( B_{y}z-B_{z}y,0,0\right) $, where $\mathbf{B}=\nabla \times \mathbf{A}$. The Hamiltonian reads
\begin{equation}
H=%
\begin{pmatrix}
h_{0}\left( \mathbf{k}\right) +\Delta _{z} & \Delta _{y} \\
\Delta _{y}^{\dag } & h_{0}^{\ast }\left( -\mathbf{k}\right) -\Delta _{z}%
\end{pmatrix},
\label{Hamiltonian}
\end{equation}%
where $h_{0}\left( \mathbf{k}\right)$  $=\varepsilon_0(\mathbf{k}) \sigma _{0}+M(\mathbf{k})\sigma _{z}+A\left(\bar{k}_x C_\alpha- k_y S_\alpha \right)\sigma_x-A\left(\bar{k}_x S_\alpha+k_y C_\alpha\right)\sigma_y$
is for one of the Weyl semimetals with $  \varepsilon_0(\mathbf{k})=C_0+C_1k_z^2+C_2(\bar{k}_x^2+k_y^2)$,  $ M(\mathbf{k})=M_0+M_1k_z^2+M_2(\bar{k}_x^2+k_y^2)$. $\bar{k}_{x}=k_{x}-eA_x/\hbar$, $S_\alpha=\sin\alpha$, $C_\alpha=\cos\alpha$, and $\alpha=-\pi/4$.
$\Delta _{y}=-i e^{-i \alpha}B_{y}\mu _{B}G/4 $ and $\Delta _{z} =B_{z}\mu _{B}G/4$ are the Zeeman energies of the magnetic field.  $G=g_{s}\left( \sigma _{0}+\sigma _{z}\right) +g_{p}\left( \sigma _{0}-\sigma
_{z}\right) $, $\mu_B$ is the Bohr magneton, and $g_s=18.6$ and $g_p=2$ are the $g$ factors~\cite{Jeon14natmat}. We adopt the parameters for Cd$_3$As$_2$ as $C_0=-0.0145$ eV, $C_1=10.59$ eV\AA$^2$, $C_2=11.5$ eV\AA$^2$, $M_0=0.0205$ eV, $M_1=-18.77$ eV\AA$^2$, $M_2=-13.5$ eV\AA$^2$, and $A=0.889$ eV\AA~\cite{Cano17prbrc} and discretize the effective Hamiltonian on a 3D simple cubic lattice with a lattice constant  $a$ (Supplemental Material~\cite{Supp}). The hinge states can emerge when the magnetic field is perpendicular to direction of the Weyl nodes. For Cd$_3$As$_2$, the Weyl nodes are along the [001] ($z$) direction, so the slab can be grown along an arbitrary superposition of the [100] and [010] ($x$ and $y$) directions.

{\color{red}\emph{One-sided hinge states in energy spectra and wave function distributions}} - Figure~\ref{fig_spectrum} shows the numerically calculated energy spectra and wave function distributions in magnetic fields for a ribbon of Cd$_3$As$_2$ , with a thickness of $100$ nm, width of $200$ nm, and periodic boundary condition along the ribbon.
$B_y$ is fixed at $15$ T.
Three cases are considered, with the tilting angles $\theta=0$ and $\theta=\pm 5^\circ$ ($B_z=B_y \tan\theta$). For each case, the energy spectrum opens a gap near $E=0$. When the magnetic field is applied along the $y$ direction, i.e., $\theta=0$,  there appear hinge states [light purple points in Fig.~\ref{fig_spectrum}(a$_1$)] on four hinges of the slab, which move along the opposite directions on the opposite side surfaces [Figs.~\ref{fig_spectrum}(a$_{2}$), \ref{fig_spectrum}(a$_{3}$), \ref{fig_spectrum}(a$_6$)]. Also, the gray points correspond to the states composed of the Landau levels on the top and bottom surfaces and the chiral states on the side surfaces [Figs.~\ref{fig_spectrum}(a$_4$) and \ref{fig_spectrum}(a$_5$)].
The side-surface chiral states arise from the
quantum Hall effect of the confinement-induced bulk-state subbands.

For the case of a tilted magnetic field ($\theta=5^\circ$), the hinge states become one sided when the Fermi energy is located in the Landau-level gap~[Fig.~\ref{fig_spectrum}(b$_1$)-\ref{fig_spectrum}(b$_6$)].
At $E_F=-4.5$ meV, we observe well-localized one-sided hinge states [Fig.~\ref{fig_spectrum}(b$_2$)-\ref{fig_spectrum}(b$_3$)  and \ref{fig_spectrum}(b$_6$)]. At $E_F=0$ meV, the one-sided hinge states are swapped to the opposite sides compared to those at $E_F=-4.5$ meV [Fig.~\ref{fig_spectrum}(b$_4$)-\ref{fig_spectrum}(b$_5$) and \ref{fig_spectrum}(b$_6$)]. The $E_F=0$ hinge states are not so localized, due to the coupling of the one-sided hinge states to the side-surface chiral states.
More interestingly, the one-sided hinge states in Figs.~\ref{fig_spectrum}(b$_1$)-\ref{fig_spectrum}(b$_6$) can be swapped by reversing $\theta$ from positive to negative [Figs.~\ref{fig_spectrum}(c$_1$)-\ref{fig_spectrum}(c$_6$)].

{\color{red}\emph{Hinge states as Landau-level edge states of Fermi arcs}} - The above evolution of the hinge states can be understood by the Fermi arcs, which are tunable by the Zeeman fields. The Fermi arcs are the topologically protected surface states in the Weyl semimetal.
The hinge states are the edge states of the Landau levels of the Fermi-arc surface states in magnetic fields.
The half-top-half-bottom Fermi-arc surface states in Fig.~\ref{fig_illustration}(b) can support the 3D quantum Hall effect and the related one-sided hinge states. However, in only $B_y$ ($\theta=0$), the Zeeman coupling $\Delta_y$ couples the Weyl semimetal and its time reversal in the Dirac semimetal, and as a result, the Fermi arcs are evenly distributed on top and bottom surfaces, as shown in Figs.~\ref{fig_spectrum}(a$_7$) and (a$_8$).
This gives the hinge states on both sides in Fig.~\ref{fig_spectrum}(a$_6$).
By contrast, the Zeeman splitting $\Delta_z$ along the $z$ direction can decouple
the Fermi arcs on opposite surfaces [see Figs.~\ref{fig_spectrum}(b$_7$)-\ref{fig_spectrum}(b$_8$) for $\theta=5^\circ$ and Fig.~\ref{fig_spectrum}(c$_7$) for $\theta=-5^\circ$], to yield the one-sided hinge states for $\theta=\pm 5^\circ$ (see Supplemental Material \cite{Supp}, Sec.~SI  for more details).

{\color{red}\emph{Landau-level gap that protects one-sided hinge states}} - Our calculations reveal that the Zeeman coupling has to be large enough ($B_y>10$ T and $\theta>5^\circ$)
to induce a Landau-level gap large enough to protect the one-sided hinge states~(more details in Supplemental Material~\cite{Supp}, Sec.~SII).
To explain the Landau level gaps in Fig.~\ref{fig_spectrum}, we numerically calculate the energies of the Landau bands of the Dirac semimetal in magnetic fields (Supplemental Material~\cite{Supp}). The external magnetic field has three effects, the Zeeman coupling ($\Delta_y$) and splitting ($\Delta_z$) effects and the orbital (Landau levels) effect [Eq.~\eqref{Hamiltonian}]. First, we consider that there are no Zeeman effects, i.e., $\Delta_y=\Delta_z=0$. As shown by the energy spectra in Figs.~\ref{fig_Landaulevel}(a) and (d), the Landau bands are doubly degenerate because the Dirac semimetal consists of two Weyl semimetals (see Supplemental Material~\cite{Supp}, Sec.~SI for more details).
Second, as shown in Figs.~\ref{fig_Landaulevel}(b) and \ref{fig_Landaulevel}(e), a nonzero Zeeman coupling $\Delta_y\neq0$ couples the two Weyl semimetals in~Eq.~\eqref{Hamiltonian} and opens the gap. The size of the gap $E_g$ increases linearly with the magnitude of the magnetic field [Fig.~\ref{fig_Landaulevel}(e)].
Finally, the Zeeman splitting $\Delta_z\neq0$ in the tilted magnetic field lifts the twofold degeneracy of the energy spectra, and then the one-sided hinge states could appear in the Landau-level gap.

{\color{red}\emph{Stability of one-sided hinge states against disorder}} - Now we check the stability of the one-sided hinge states against disorder,
by calculating the conductance using the
Landauer-B\"uttiker formula
\cite{Landauer1970Philosophical,Buttiker88prb,
Fisher1981PRB} and the recursive Green's
function method \cite{Mackinnon1985Zeitschrift,Metalidis2005PRB} (Supplemental Material~\cite{Supp}, Sec.~SIII).
We adopt the Anderson-type disorder
by considering random on-site energies fluctuating in the energy interval $[-U,U]$, where $U$ is the disorder strength. Figure~\ref{fig_disorder}(a) shows the conductance as a function of the disorder strength $U$ for $E_F=-4.5$ meV and $\theta=5^\circ$. In the clean limit, the system is a quantum Hall insulator with a quantized conductance $G=e^2/h$. The quantized conductance is from the one-sided hinge states, as confirmed by Fig.~\ref{fig_disorder}(b),
which shows the disorder-averaged local density of states (Supplemental Material~\cite{Supp}, Sec.~SIII).
With increasing disorder strength, the conductance remains quantized until the disorder strength $U$ exceeds $40$ meV, which is much larger compared to the Landau-level gap (about 10 meV). Therefore, we show that the one-sided hinge states are robust against weak disorder. With a further increase in the disorder strength, the conductance drops, but with nonvanishing fluctuations, indicating a signature of the diffusive metal phase~\cite{ChenCZ15prl,Chen2018PRBWeyl}. The diffusive metal phase is generally observed in 3D disordered systems, which further confirms the 3D nature of the Dirac semimetal ribbon that supports the 1D one-sided hinge states. Moreover, the quantized conductance is characterized by a nonzero Chern number $C=1$ (see Supplemental Material~\cite{Supp}, Sec.~SII for more details).

{\color{red}\emph{Scanning tunneling Hall measurement of one-sided hinge states}} -
The one-sided hinge states provide a platform for experimental observation of the Fermi-arc 3D quantum Hall effect and higher-order topological states of matter. Naively thinking, the one-sided hinge states can be probed with the help of the scanning tunneling Hall measurements shown in Fig.~\ref{fig_disorder}(c), which combines the capabilities of the scanning tunneling microscopy (STM) and the in-plane magnetotransport~\cite{Schwenk2020RSI}.
The STM tip can probe the local density of states $\rho\left(E,y,z\right)=\int_{k_x}\sum_n \left| \psi_n\left(k_x,y,z\right) \right|^2\delta\left(E-E_n\right)$ of the hinge states.
Figure~\ref{fig_disorder}(d) shows the local density of states of the topmost layer, i.e., $\rho\left(y=n_y,z\right)$,
as a function of $z$. For $\theta=0$, the two hinges of the topmost layer have the same spectral weight due to the degeneracy of the two Weyl semimetals when $\Delta_z=0$. In Fig.~\ref{fig_disorder}(d$_3$), the nonvanishing local density of states in the central region is attributed to the Landau levels, which contribute to neither the in-plane transport nor the local density of states at the two hinges. Interestingly, for $\theta=5^\circ$, the local density of states becomes zero on one side when $E_F=-4.5$ meV~[Fig.~\ref{fig_disorder}(d$_2$)], and is nonzero on both sides when $E_F=4.5$ meV~[Fig.~\ref{fig_disorder}(d$_4$)].

Considering the hinge states penetrate up to 20 nm in Fig.~\ref{fig_disorder}(d), we define $D_L=\sum_{z=1}^{20}\rho\left(n_y,z\right)$ and $D_R=\sum_{z=181}^{200}\rho\left(n_y,z\right)$ as the total local density of states for the outermost 20 layers near the two hinges. As shown in Figs.~\ref{fig_disorder}(e)-\ref{fig_disorder}(f), for $E_F=-4.5$ meV, $D_R$ vanishes when $\theta>3^\circ$, due to the decoupling of $h(\mathbf{k})$ and $h^*(-\mathbf{k})$. Such a behavior is absent for $E_F=4.5$ meV, at which both $D_L$ and $D_R$ oscillate, due to the band crossings in the energy spectra. The above behaviors can be seen more clearly in the phase diagram shown in Fig.~\ref{fig_disorder}(g). The red and blue patches indicate the phase of the one-sided hinge states, and they are swapped as $E_F$ and $\theta$ changes, corresponding to from Figs.~\ref{fig_spectrum}(b$_6$) and \ref{fig_spectrum}(c$_6$).

Furthermore, we expect that the one-sided hinge states can also be probed by using the Aharonov-Bohm effect, e.g., in a Fabry-P\'{e}rot interferometer~\cite{li2021higher}.

\begin{acknowledgments}
This work was supported by the National Natural Science Foundation of China (Grants No. 11925402 and No. 11974249), the National Basic Research Program of China (Grant No. 2015CB921102), the Strategic Priority Research Program of Chinese Academy of Sciences (Grant No. XDB28000000), the Natural Science Foundation of Shanghai (Grant No. 19ZR1437300), Guangdong province (Grants No. 2020KCXTD001 and No. 2016ZT06D348), Shenzhen High-level Special Fund (Grants No. G02206304 and No. G02206404), and the Science, Technology and Innovation Commission of Shenzhen Municipality (Grants No. ZDSYS20170303165926217, No. JCYJ20170412152620376, and No. KYTDPT20181011104202253). R. C. acknowledges support from the project funded by the China Postdoctoral Science Foundation (Grant No. 2019M661678) and the SUSTech Presidential Postdoctoral Fellowship. The numerical calculations were supported by Center for Computational Science and Engineering of SUSTech.
\end{acknowledgments}

\bibliographystyle{apsrev4-1-etal-title_6authors}
\bibliography{bibfile,refs-transport}


%
\end{document}